\documentclass[preprint]{article} 
\usepackage{natbib}
\usepackage{graphicx}
\usepackage{amssymb}
\begin{document}

\message{Ordinary=fast axis; Extraordinary=slow axis; optic axis defined by orthogonal direction}

\message{Correlation with dust-logger}

{\large\centerline{Radio Frequency Birefringence in South Polar Ice and Implications for Neutrino Reconstruction}}

\begin{center}
I. Kravchenko \\
{\it University of Nebraska, Department of Physics and Astronomy, Lincoln, NE, 68588-0111} \\
D. Besson, A. Ramos, J. Remmers \\
{\it University of Kansas, Dept. of Physics and Astronomy, Lawrence, KS, 66045-7582} \\
\end{center}

\vspace{1cm}
\begin{abstract}Using a bistatic radar echo sounding (RES) system developed for
calibration of the RICE particle astrophysics
experiment at the South Pole, 
we have studied radio frequency (RF) reflections off the bedrock. 
The total propagation time of $\sim$ns-duration, vertically (${\hat z}$) broadcast radio signals, 
as a function of polarization orientation in the horizontal plane, 
provides a direct probe of the geometry-dependence of the ice 
permittivity to a depth of 2.8 km. 
We observe clear birefringent asymmetries along ${\hat z}$
in the lowest half of the ice sheet, at a fractional level $\sim$0.3\%. 
This result is in contrast to expectations based on measurements at Dome Fuji, for which birefringence was observed in 
the upper 1.5 km of the ice sheet.
This effect, combined with the increased radio frequency attenuation expected near the bedrock, 
renders the lower half thickness of South Polar ice less favorable than the
upper half of the ice sheet in terms 
of its ultra-high energy neutrino detection potential.
\end{abstract}


\section*{Introduction}
The response of ice as a function of polarization (``birefringence'') is characterized
by differences in either wavespeed or absorption along linear
(generally orthogonal) axes. 
Over a km-scale pathlength,
in the absence of any preferred in-ice direction, one might
expect any time propagation
asymmetry at the single-crystal level
to be macroscopically mitigated by the randomness of 
the corresponding single-crystal
orientation. In such a case, over a total pathlength $l$ consisting
of $N$ unit steps, each 
characterized by an asymmetry $b$, the average
propagation time along each polarization axis should have a Gaussian
distribution, centered at $l/c$, with width $b\sqrt{N}l/Nc$. The asymmetry
distribution would therefore be a Gaussian of width $\sigma_b=b\sqrt{2N}l/Nc$,
centered at zero.
For 1\% birefringence ($b=0.01$), $l$=1000 m, and
step sizes corresponding to 
typical grain sizes ($10^{-3}$ m, or N=$10^6$),\message{1 cm in GRIP core - see Danish thesis}
we expect $\sigma_b\lesssim$0.1 ns. 
If, however, 
bulk flow of the ice sheet results in a preferred in-ice direction much longer
than typical grain sizes, the 
propagation time asymmetry can be ${\cal O}$(10 ns).

\subsection*{Impact on Particle Astrophysics}
Efforts are underway to use the Antarctic ice sheet as an ultra-high
energy neutrino
target\citep{ICECUBE,RICE,ANITA}. 
Neutrino-ice
collisions result in the production of 
charged particles which
emanate from the interaction point with velocities approaching
the speed-of-light {\it in vacuo}: $v\to c$.
In a medium with index-of-refraction $n>1$, detection of the resulting
Cherenkov radiation
in either the near-UV (TeV-scale neutrinos) 
or radio wavelength regime (PeV-scale neutrinos) by a suite of
sensors can be used to reconstruct the kinematics of the initial
neutrino, provided the 
absorption and refraction of the original electromagnetic
signal due to the intervening ice can be reliably 
estimated. 
The RICE experiment\citep{RICE} demonstrated the feasibility of
the radio-detection approach over the last decade.
The Askaryan Radio Array (ARA) Collaboration\citep{ARA}
seeks to substantially enlarge the current
RICE footprint at South Pole
by instrumenting 
an 80~$km^2$ area over the period 2010-2015. 
Ice birefringence could result in an initially ns-scale RF pulse 
being resolved into
two components, with a time stagger comparable to the signal
duration itself, requiring a trigger system with a correspondingly long
signal integration time.
Complete characterization of the ice permittivity,
as a function of depth, and also polarization is
therefore essential in obtaining a reliable estimate of the
neutrino detection efficiency. 

\section*{Birefringence Measurements}
\subsection*{Prior Direct Experimental Work}
Dielectric studies of ice have
been ongoing for nearly a century\citep{Ehringhaus17},
with radio frequency studies well over half a century old\citep{AutyCole52}.
The study of ice birefringence is itself a discipline which is
also several decades old\citep{Hargreaves-1978}. At visible wavelengths (590 nm),
ice exhibits a birefringent asymmetry of $\sim$0.3\%. 
Early RF studies focused on single-crystal measurements in the lab,
with complementary samples of data collected {\it in situ}.
Even prior to the advent of ns-scale signal capture, 
the field was already quite mature, with estimates of birefringent
asymmetries of order 1\%\citep{Hargreaves-1977}.

To explain the asymmetry observed
in those early {\it in situ} data,
the possible effects of geometrically distorted air
bubbles and/or other sources of in-ice density contrast, internal layering,
and the inherent single-crystal anisotropy, macroscopically
manifest as a preferred axis of the crystal orientation fabric (COF),
were all considered. 
The physical
mechanism for COF alignment, however, was not fully explained --
experimental data up to that point contradicted the hypothesis that internal
strain due to ice flow is exclusively responsible for the observed birefringence.

Briefly, crystal orientation correlates with depth as follows\citep{Beau08}: 
i) In deeper Antarctic ice, crystals align with the 
horizontal
glide plane; near the surface, there is no preferred orientation, ii) 
complete COF
alignment results in an order-of-magnitude faster bulk flow than randomly oriented COF.
Note the interplay between ice flow and COF alignment - as shear increases, COF alignment increases, which, in turn, facilitates ice flow. The effect can be particularly dramatic at ice divides\citep{Martin2008}, given the large internal strains expected at those locales.
The correlation between COF alignment and ice flow was
verified by an analysis of airborne radar (60 MHz) data taken in the vicinity
of Vostok Station, where the derived 
East Antarctic COF alignment was found to corroborate direct Vostok ice core data\citep{Siegert00}.
That study produced many results relevant to our investigation -- most notably the conclusion that
the variation of echo strength with increasing
depth indicates scattering dominated by density, acidity, and COF effects, in the top, middle,
and bottom third of the ice sheet at Vostok. To the extent that birefringence results from
COF, we would therefore expect that non-zero birefringence would be most noticeable for 
deeper returns.
Similarly, airborne-sounding data taken at Dronning Maud Land, some 3000 km West of Vostok, were also compared against ice-core data\citep{Eisen07}. 
150 MHz pulses of 60 ns and 600 ns duration revealed the presence of an horizontally extended ($\sim$5 km) plane, at a depth of 2025--2045 m, or approximately 800 m above the bed. That deep 
reflector was found to be inconsistent with
acidity scattering, and more compatible with a discontinuity in the COF.
Direct comparison of Vostok, or Dronning Maud data 
to South Pole, however, must properly
take into account the variation in ice flow due to the 
different underlying stratigraphy and the fact
that much of the ice sheet at Vostok lies in proximity to
a nearly resistance-less lake.

\begin{figure}[htpb]
\begin{minipage}{18pc}
\centerline{\includegraphics[width=7cm]{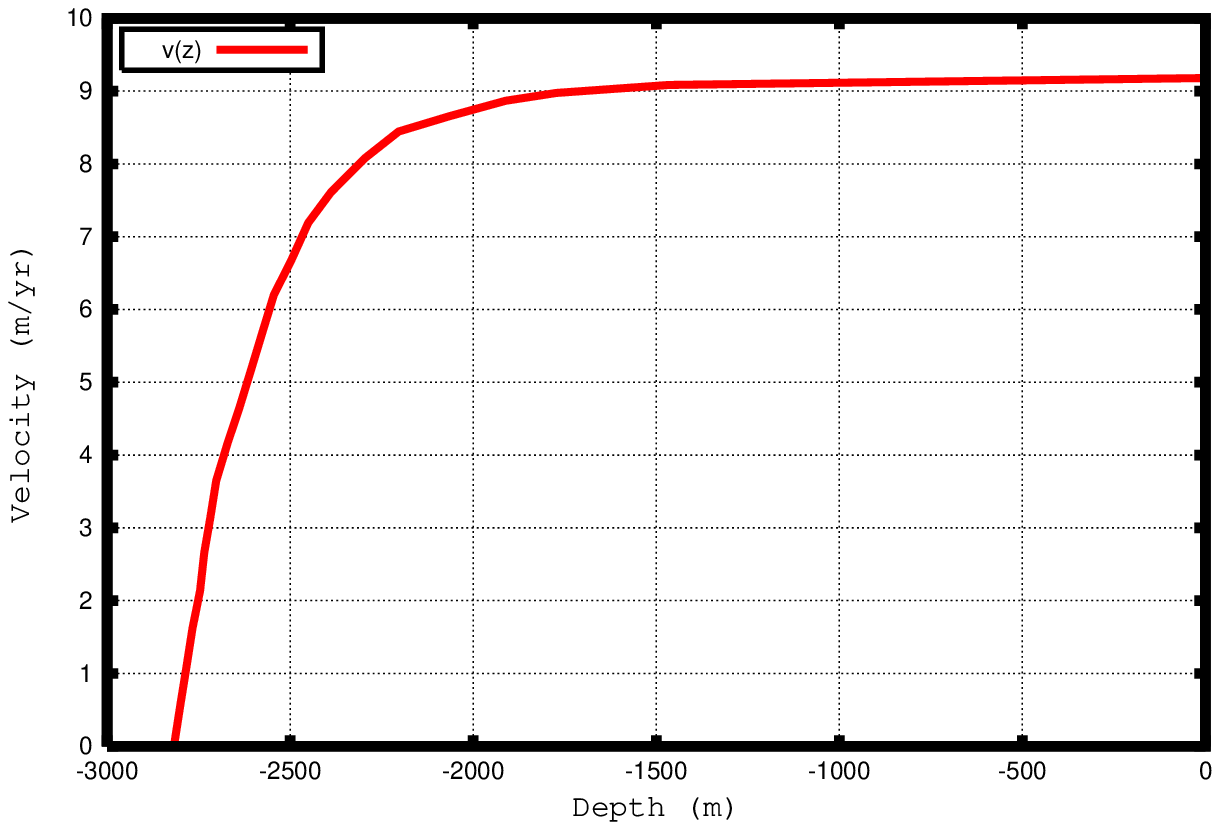}}
\caption{Extracted ice flow velocity profile vs. depth at South Pole, taken from \citep{ice-shear}.}
\label{fig:iceflow-v-depth.eps}
\end{minipage}
\hspace{1pc}
\begin{minipage}{18pc}
\centerline{\includegraphics[width=8cm]{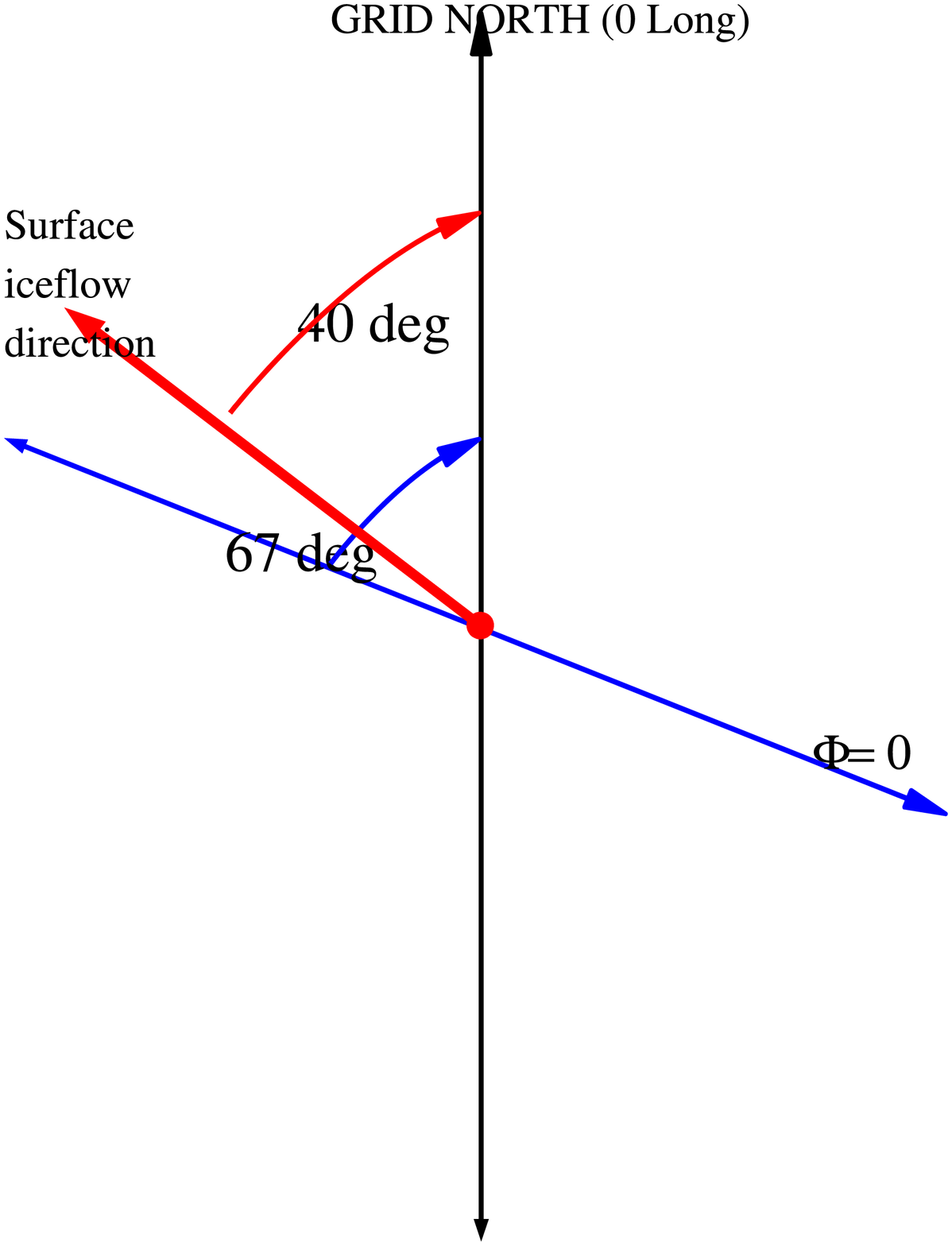}}
\caption{Geometry of measurements presented herein. $\phi$=0 corresponds
to the reference polarization orientation for TEM horns, chosen to correspond to
the long axis of the MAPO building for convenience.
Surface ice flow direction at South Pole is shown in red, corresponding
to $\phi$=--27$^\circ$ ($\equiv$+153$^\circ$). Magnitude of surface ice flow
is measured to be $\sim$9 m/yr (Fig. \ref{fig:iceflow-v-depth.eps}), and uniform to a depth of $\sim$2 km. 
(Data courtesy of Kurtis Skoog.)}
\label{fig:iceflow}
\end{minipage}
\end{figure}

Over a length scale of tens of kilometers, South Polar ice is observed to flow at 
the surface with a velocity of approximately
9--10 m/yr\citep{ice-shear} in a direction which corresponds to about 153$^\circ$ in the
coordinate system used for these measurements
(corresponding to the 40 degree West Longitude line).
The velocity profile, as a function of depth, has been determined by 
Price {\it et al.}\citep{ice-shear}, and is reproduced in Figure
\ref{fig:iceflow-v-depth.eps}.
To the extent that ice flow shear implies COF-alignment, 
and COF-alignment is singly responsible for birefringence, we would therefore expect very little birefringence to a depth of $\sim$2 km, with
increasing birefringence at greater depths, if the ice flow at present
is the same as the ice flow through the history of the ice sheet formation.



There have been several
more recent\citep{Matsuoka-biref,Matsuoka2004} {\it in situ} 
measurements of Antarctic ice\citep{Doake2002,Doake2003} which have also
been interpreted as 
evidence for COF-induced
birefringence.
In a study
of internal layer reflections
conducted at Dome Fuji\citep{Fujita-1993},
co-polarized (transmitter [Tx] and receiver [Rx] antenna
polarizations parallel, projected onto the
horizontal plane) 
and cross-polarized 
(Tx and Rx perpendicular)
measurements were made over the full azimuth. 
Signals broadcast at a given
frequency using three-element
Yagis were detected after internal scattering from within the ice;
both $\pi$ and
$\pi/2$ modulations of the received
amplitude were observed. Under the assumption that the ordinary and
extra-ordinary birefringent axes are orthogonal,
the latter was interpreted as due to
birefringent-induced interference effects in the time-domain.
In particular, as the transmitter-receiver systems
were rotated in azimuth, signal drops were observed at angles $(2m+1)\pi/4$
(with m an integer)
relative to the presumed horizontal COF-alignment, which
were attributed to destructive interference between the ordinary vs.
extra-ordinary signals at the broadcast wavelength. 
Those data\citep{Matsuoka-biref} imply $\sim\lambda/2$ variation over a
depth of 1500 meters, which (given that $\lambda_{ice}\sim$3~m for 179 MHz in-ice broadcasts)
imply a birefringent asymmetry of order 0.1\%.
The authors note that their measured
birefringence is somewhat weaker at 60 MHz, although an
authoritative recent study of pure laboratory
ice obtained an asymmetry $\delta_{\epsilon'}$
of 1.07$\pm$0.23\%\citep{Matsuoka97} at both 1 MHz and 39 GHz. 

In a comprehensive attempt to model the Dome Fuji data,
and neglecting any possible tilting of the reflecting internal layers (which 
must occur coherently
over an aeral scale of order the Fresnel zone to be significant),
the dependence of radar scattering
on density, COF, and acidity effects were assessed\citep{Matsuoka2004}.
In principle, the type of scattering can be elucidated on the basis of
signal strength:
COF and acidity-based layers typically reflect -30 -- -60 dB of the 
incident power; density scattering, integrated through a vertical chord is typically
a factor 10 larger (in power). To the extent that acidity scattering is a pure
conductivity effect, we would expect it to vary as the inverse of frequency.
To the extent that density and COF scattering is due to a variation
of $n(\omega)$, we would expect such scattering to show 
much weaker frequency
dependence, depending on the proximity of Debye resonances.
\message{Of the three, only acidity scattering is
expected to vary ($\propto$1/f)
with frequency in the RF regime, since acidity is a true conductivity effect rather
than a variation in the permittivity.}
In order to simplify the interpretation of data, the authors
made the assumption that one scattering
effect was dominant, and that
although all types of scattering can result
in large cross-polarized signals, only COF produces anisotropic scattering
which also gave the observed azimuthal
interference patterns. 



\message{Not in Table 1, but quote - This is because the effect of PCOF
appears in one orientation but the effects of the other mechanisms
can appear in the orthogonal orientation.}



In 2006, our group previously used time-domain
bedrock reflections observed at a site
near Taylor Dome\citep{TD} 
to estimate a birefringent asymmetry of 0.12\%, projected
onto the vertical ${\hat z}$-axis 
(perpendicular to the surface).
A lack of conclusive ice flow data at Taylor Dome
prevented a
correlation with the local ice flow 
direction from being established.
A follow-up study in 2008\citep{SPRefl08} at the South Pole searched for,
but found no detectable azimuthal variation in echo return times
through the upper 1600 m of the ice sheet.


Table \ref{tab:birefsum} summarizes some recent birefringence measurements.
\message{compared with a value of 2 percent for single ice crystals - reference?}
\begin{table}[htpb]
\caption{Summary of recent birefringence measurements. \label{tab:birefsum}}
\begin{center}
\begin{tabular}{c|c|c|c} 
Group & Locale & $\delta_{\epsilon'}$ Result & Comment \\ \hline
\citep{Hargreaves-1977} & Greenland & 0.024--0.031\% & \\
\citep{Doake2002} & Brunt Ice Shelf & $>$0.14--0.47\% & \\
\citep{Doake2003} & George VI Ice Shelf & $>$0.05--0.15\% & \\
\citep{Matsuoka97} & Lab Ice & $\sim$3.4\% & 1 MHz -- 39 GHz \\
\citep{Fujita-1996} & Lab Ice & (3.7$\pm$0.6)\% & 9.7 GHz \\ 
\citep{WoodruffDoake79} & Bach Ice Shelf & 0.52\% & \\
\citep{Fujita-2003} & Mizuho Station & measurable & \\ 
\citep{Fujita06} & Mizuho & 1.5\%-3.5\% & frequency-domain \\
\citep{TD} & Taylor Dome & 0.12\% & time-domain \\ \hline
\end{tabular}
\end{center}
\end{table}

\begin{figure}[htpb]
\centerline{\includegraphics[width=9cm]{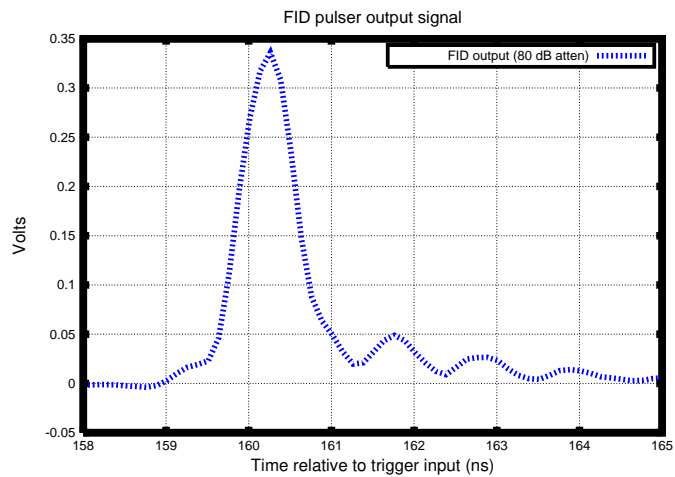}}
\caption{Output signal from FID pulser, attenuated by 80 dB.}
\label{FID-output}
\end{figure}

\section*{Experimental Configuration}
The Martin A. Pomerantz Observatory (MAPO) building, located at the South Pole, is used to house
the signal generator and data acquisition system used for our measurements.
Two 1.25-cm thick LMR-500 coaxial cables, 
each approximately 40 m long, were fed 
from within MAPO
through a
conduit at the bottom of the building and out onto the snow, and connect to the
transmitter and receiver horn antennas (separated by 50 m, and on opposite sides of the
MAPO building), respectively.
Complete details of this 
measurement, including calibration of
antennas, amplifiers, filters, cables, and
verification of both the thermal noise floor, as well
as the determination of the polarization dependence of
the output signal, 
are provided in our 2008 study, probing the
upper half of the ice sheet\citep{SPRefl08}. 
Our experimental geometry, relative to the surface ice flow axis, is 
identical to those previous measurements, and shown in
Figure \ref{fig:iceflow}. 
As before, a LeCroy 950 Waverunner
digital oscilloscope is used for data capture.
The primary difference is that, for the measurements
described herein, higher-power pulsers are used, with a maximum 
output voltage of $\sim$3000 V. 
These higher-power pulsers now permit observation of the
bottom reflection itself. 
To ensure consistent results, we employed two pulsers (Grant Corporation
HYPS, and FID model FPG6-1PNK), two
sets of antennas (a pair of TEM horn antennas built by the Institute of
Nuclear Resarch [INR], Moscow, Russia and two 
Seavey Co. quad-ridge, dual-polarization antennas of the
same type used by the ANITA experiment), and also two amplifier gains (36 dB, or 52 dB gain), with
a variety of high-pass filters to allow multiple comparisons of signal shape and amplitude.
The output signal shape from the FID pulser is shown in Figure \ref{FID-output}.
The combination of the FID pulser plus the INR horn antennas, at 36 dB gain, with 
100 MHz high-pass filtering 
were found to yield the best signal-to-noise, without amplifier saturation.
This configuration provides the bulk of the results herein, and is defined as
the `Run 1' configuration. To improve signal:noise, waveforms are typically averaged over $\sim$10000 captures.
Each averaged
waveform is designated by the angle of the transmission polarization
axis relative to our zero-degree reference (Fig. \ref{fig:iceflow}), as well as the 
corresponding angle for the receiver axis.


\section*{Observed Signals}
We attempted to detect birefringent effects using the same approach as
that used in our previous Taylor Dome
analysis\citep{TD}; namely, we search for
a measurable time difference in received signals, as a function of
the orientation of the long axis (transmitted signal
polarization axis) of the horn antennas.
Although the received spectral power is expected to be mostly determined by the
$\lambda^2$ dependence of the horn antenna effective area,
we have examined the signal strength as a function of frequency (transforming
32-ns [64-sample] time segments), as well, 
presuming that the frequency characteristics of the bedrock reflection may be
differentiable from internal layer scattering, density scattering, etc. and
thereby help elucidate distinct echoes.

\begin{figure}[htpb]
\centerline{\includegraphics[width=14cm]{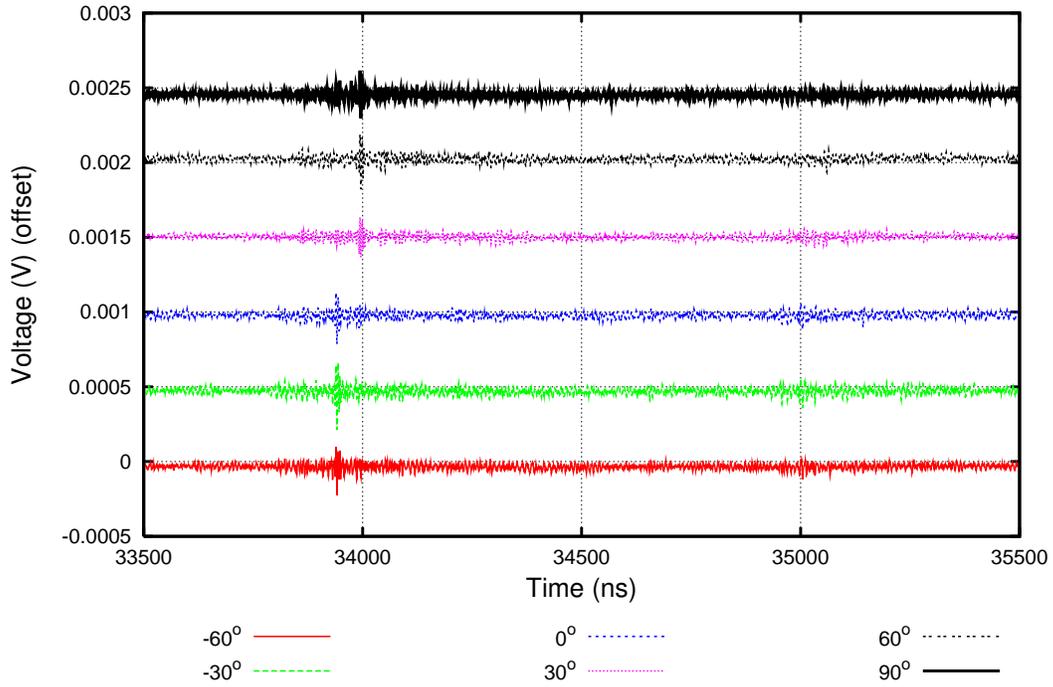}}
\caption{Measured voltages, as a function of co-polarization orientation of
transmitter and receiver, for times consistent with the expected bedrock reflection.
Waveforms have been vertically offset for clarity.}
\label{fig:RT090}
\end{figure}

\begin{figure}[htpb]
\centerline{\includegraphics[width=13cm]{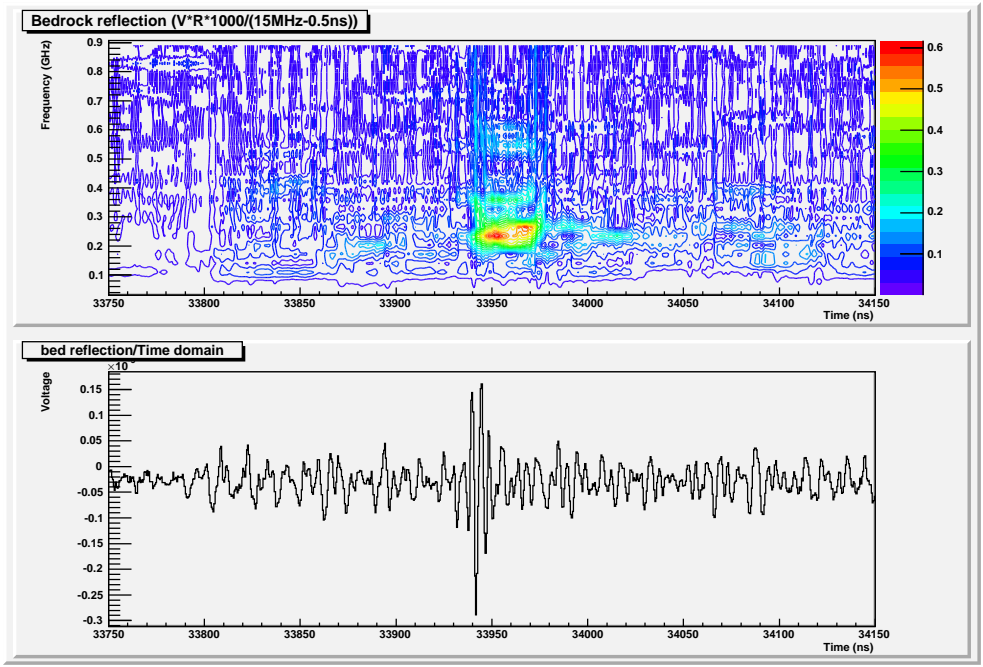}}
\caption{Power spectrum of received signal strength (relative units) for echo
return consistent with time delay expected for bedrock reflection; antennas
aligned at $-30^o$.}
\label{fig:SPRefl09}
\end{figure}

\begin{figure}[htpb]
\centerline{\includegraphics[width=13cm]{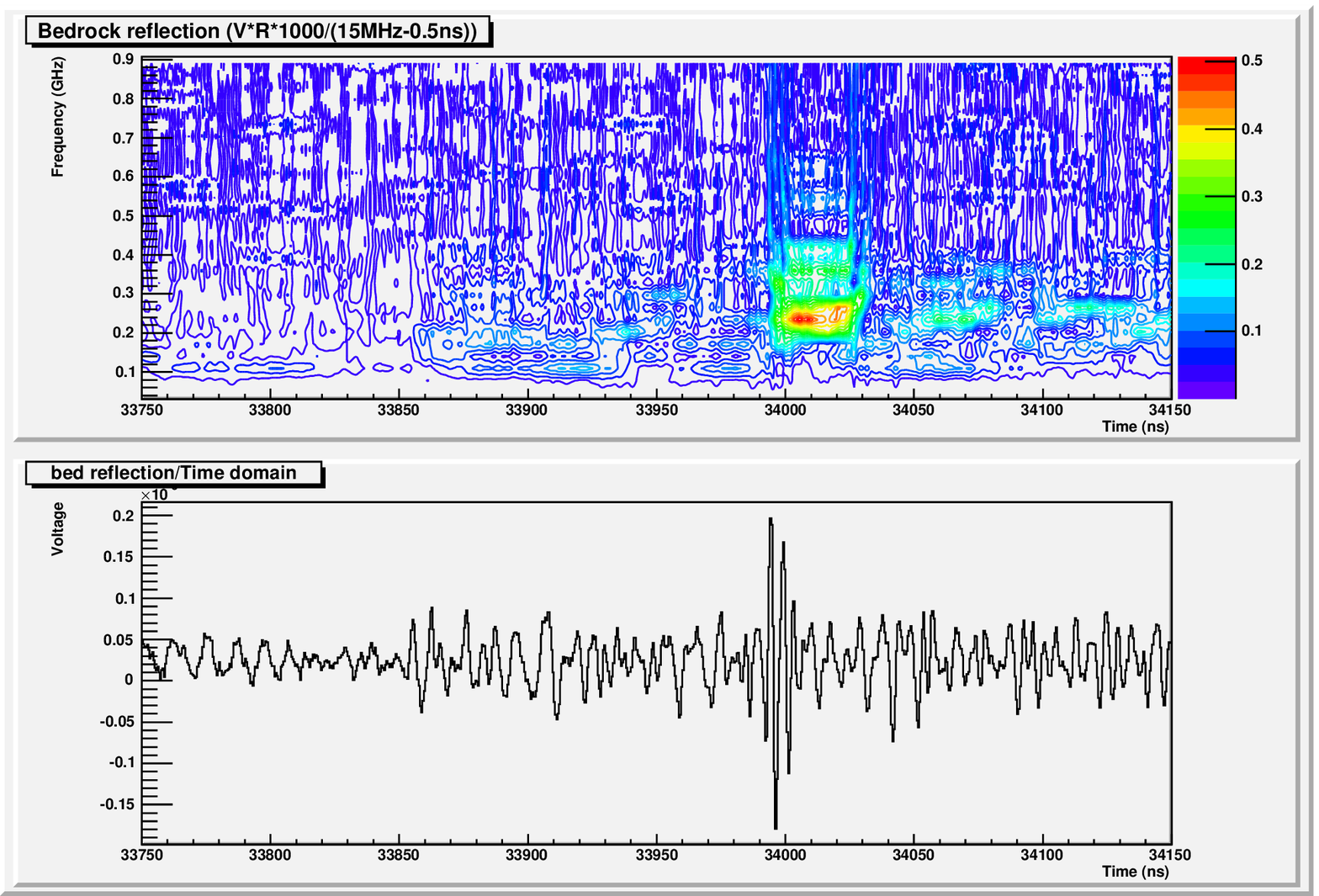}}
\caption{Power spectrum of received signal strength (relative units) for echo
return consistent with time delay expected for bedrock reflection; antennas
aligned at $+60^o$.}
\label{fig:SPRefl09_1}
\end{figure}

The obtained voltages bracketing the time for the bedrock 
reflection ($\sim$34 $\mu$s), as a function of rotation angle, are shown in Figure
\ref{fig:RT090}.
As we sweep in azimuth, we note an apparent shift in the bedrock reflection time,
between the zero degree and +30 degree orientation data sets.
Figures \ref{fig:SPRefl09} and \ref{fig:SPRefl09_1} present the 
data for the -30 degree vs. +60 degree orientations, 
zooming in around the expected bedrock reflection time to illustrate
the $\sim$50 ns time shift,
and also showing the Fourier transform (FT). 

\begin{figure}[htpb]
\centerline{\includegraphics[width=14cm]{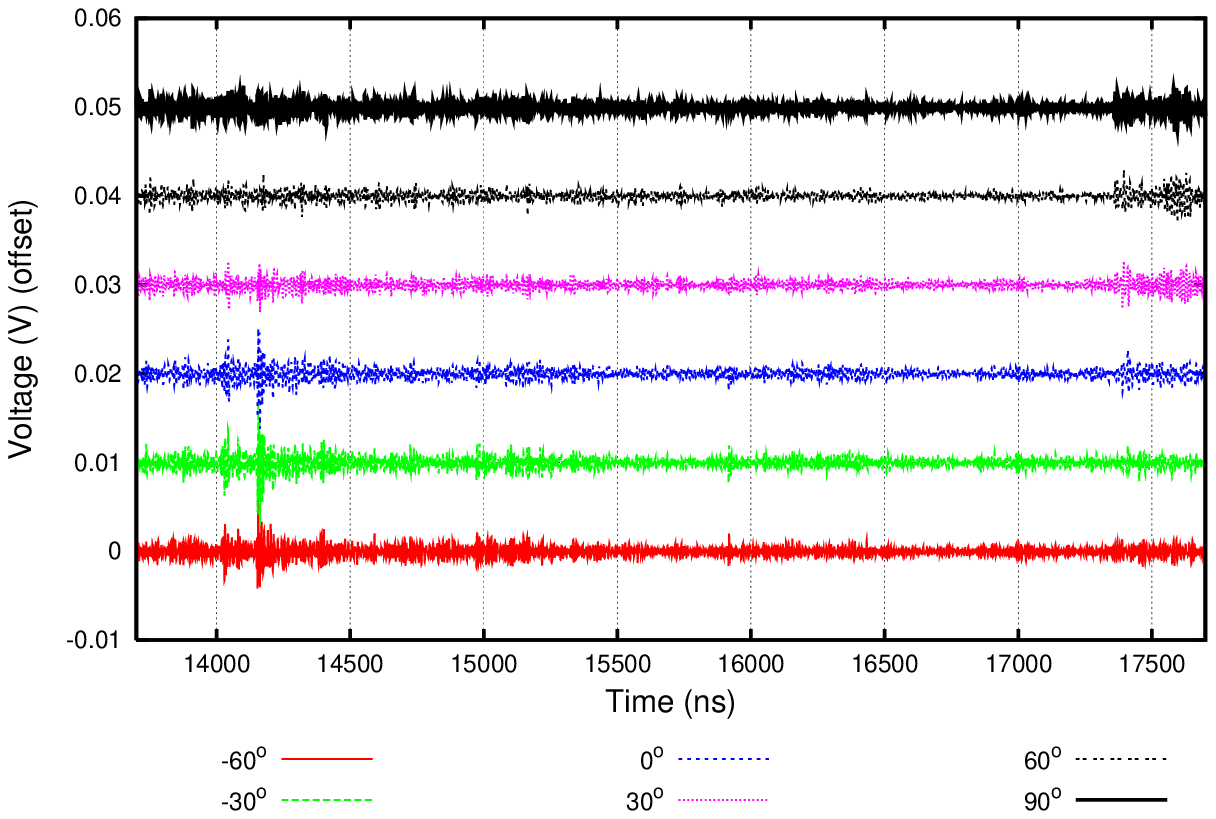}}
\caption{Measured voltages, as a function of co-polarization orientation of
transmitter and receiver, for times consistent with reflections from within the ice itself.
Upper five waveforms have been vertically offset for clarity.}
\label{fig:RT090_0}
\end{figure}

\begin{figure}[htpb]
\centerline{\includegraphics[width=14cm]{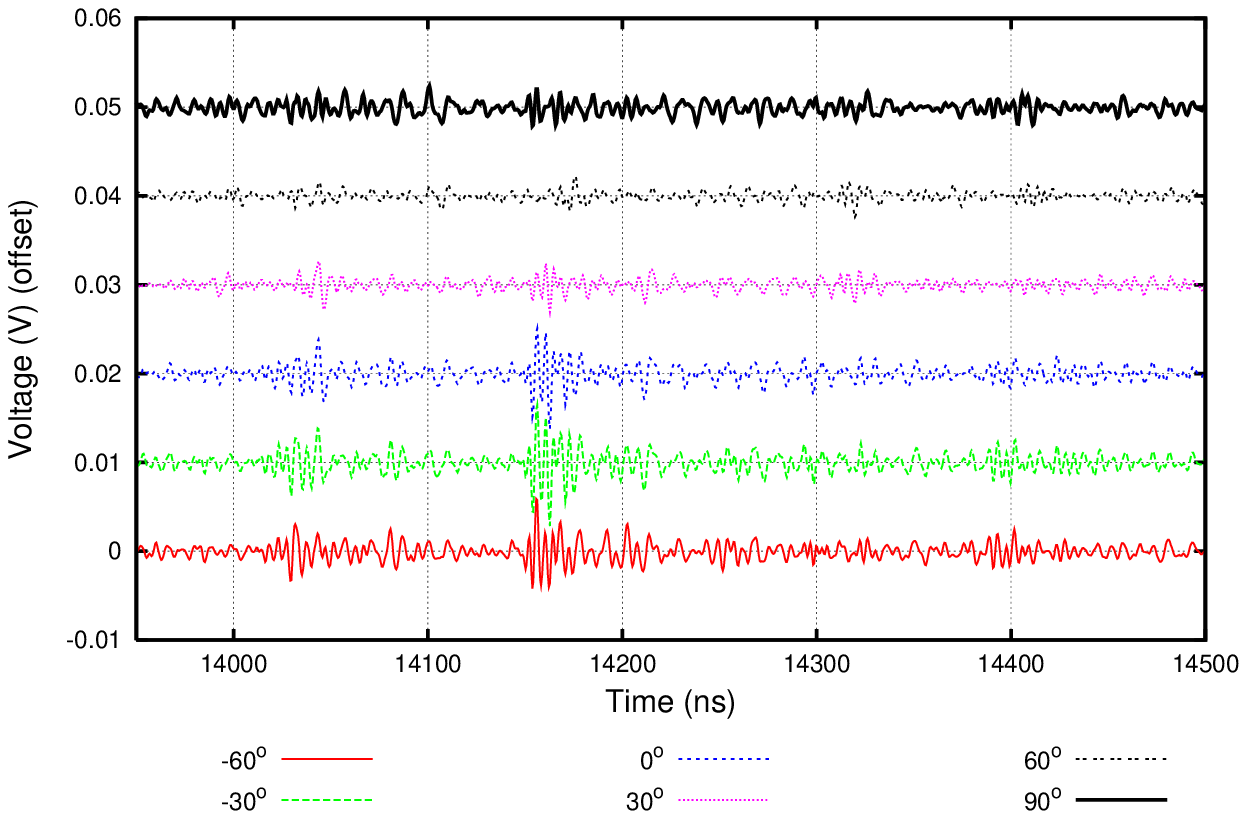}}
\caption{Zoom of previous figure.}
\label{fig:RT090_1}
\end{figure}

\begin{figure}[htpb]\centerline{\includegraphics[width=15cm]{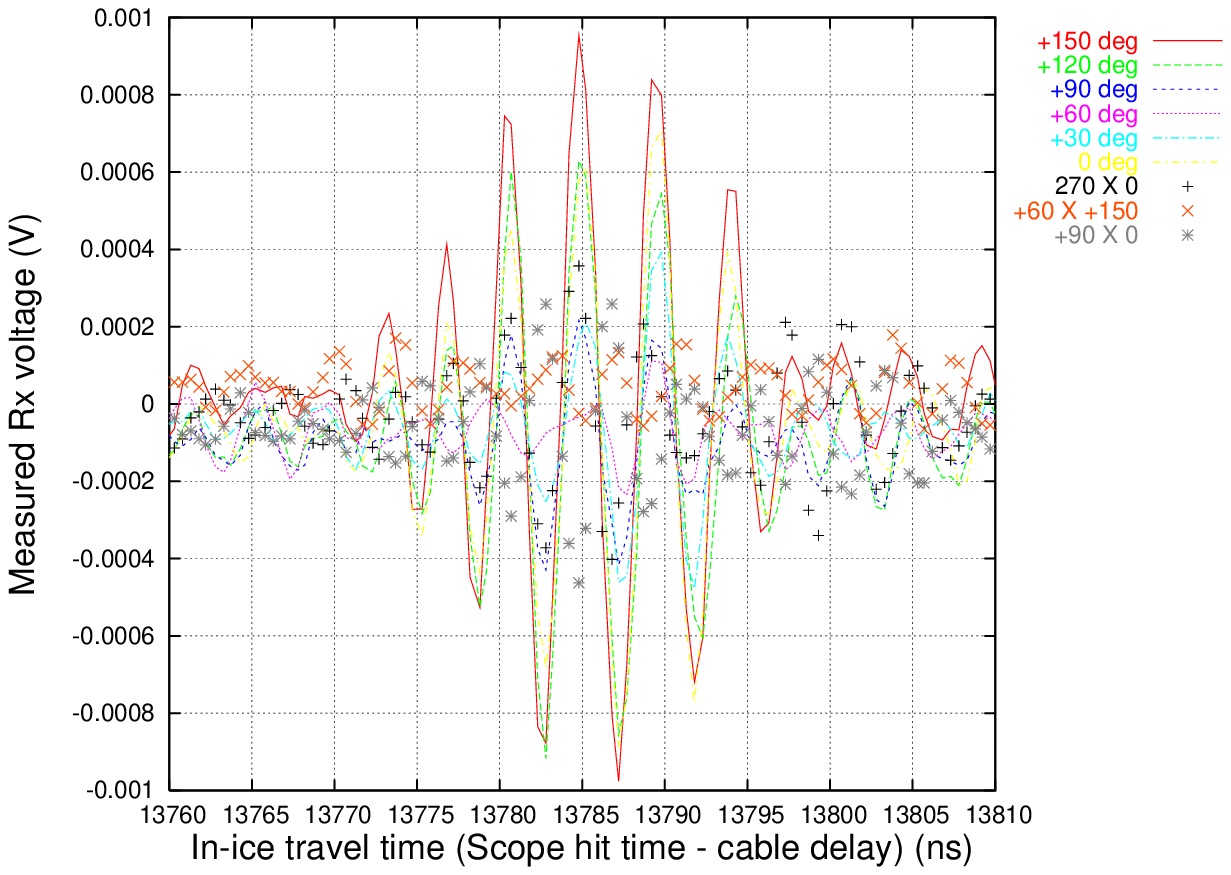}}\caption{Ensemble of reflections observed in time interval around 14 microseconds after trigger,
taken from 2008 South Polar data. 
Angular orientation convention is the same as for the current analysis.
Cross-polarized reflections are shown as points in this plot.}\label{fig:13.9us-all}\end{figure}

By comparison, the echo returns observed at earlier times are shown in
Figure \ref{fig:RT090_0} and zoomed in Figure \ref{fig:RT090_1}.
Based on the time stagger in the bedrock 
reflection, and assuming that the time delay is linear with ice depth, we would expect to observe a 
corresponding 20 ns
delay at an echo time of $\sim$14 $\mu$s. 
No such obvious
shift is observed, consistent with the results obtained in our 
previous study
(see Fig. \ref{fig:13.9us-all}, taken from our 2008 South Polar data analysis).  
Our data therefore suggest that the birefringent asymmetry is 
primarily generated in the lower half
of the ice sheet.

\begin{figure}[htpb]\centerline{\includegraphics[width=9cm]{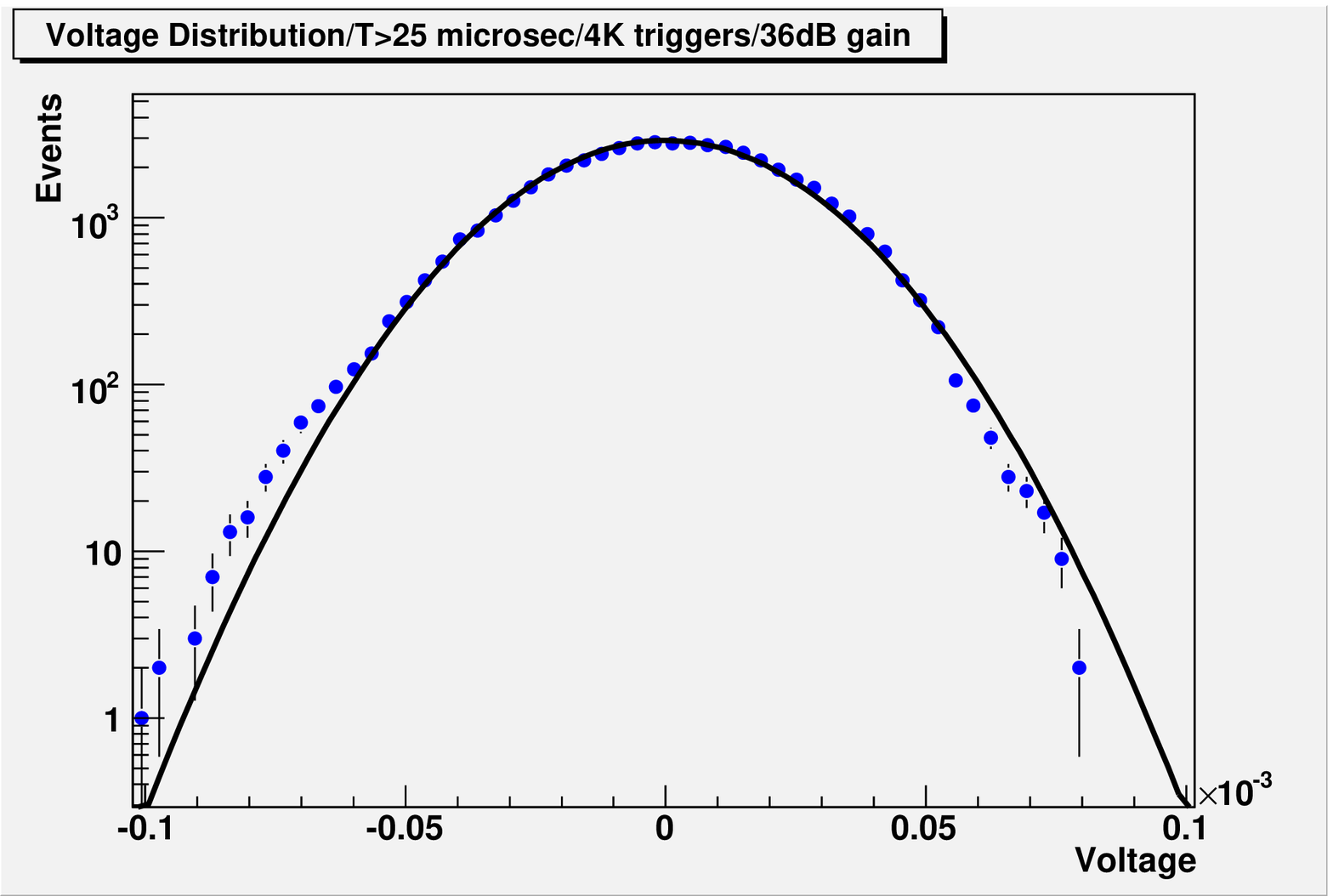}}\caption{Gaussian fit to voltage distribution showing expected thermal-like characteristics.}\label{hV_fit_log.eps}\end{figure}

Ideally, we would have sufficient resolution to observe discrete scattering layers near the bed, and thereby map the magnitude of the birefringent
delay as a function of depth. Although a visual inspection of the waveforms reveals
the presence of enhancements in the received voltage at echo times of $\sim$24 and
$\sim$26 microseconds, they are not sufficiently convincing to allow an
extraction of a birefringent asymmetry.
Previous RES measurements have, in fact, similarly observed a lack of evident
echoes in deep ice for $\sim$10 microseconds preceding the bedrock echo (the
so-called ``Echo-Free Zone'' (EFZ)\citep{Robin1977,Fujita1999,Matsuoka2003}) 
The EFZ is illustrated by the distribution of recorded voltages, corrected for averaging and amplification, for echo times greater than 25 microseconds, as shown in Figure \ref{hV_fit_log.eps}. The expected rms voltage, assuming a 220 K environment and comparable system noise, and using $V_{rms}=\sqrt{4k_BTB}$ with $k_B$ Boltzmann's constant and $B$ the bandwidth of our receiver system ($\sim$1 GHz) is approximately 20 microVolts.


\subsection*{Bedrock Echo Signal Shape}
Figures \ref{fig:SPRefl09} and \ref{fig:SPRefl09_1} display an apparently
extended signal shape, including some enhanced power apparently preceding the
primary bedrock reflection. This extended signal shape, persisting for $\sim$200--300 ns, was similarly observed in our studies at Taylor Dome, and is also present in the bedrock reflection data taken by ground-penetrating radar measurements from, e.g, the Center for Remote Sensing of Ice Sheets (CReSIS)\citep{CRESIS}.
The primary reflecting area on the bedrock surface is expected to be roughly one Fresnel zone wide, or approximately $\sqrt{2\lambda d}$ in radius. With $\lambda\sim$1 m, and $d\sim$5.6 km, the time difference between a normally incident ray and a ray reflecting at the periphery of the first Fresnel zone should be $\sim (d/c)(1+\lambda/d)$, or approximately 20 ns, considerably smaller than our observations. Given the typical antenna
beamwidth of $\sim$30 degrees in air ($\sim$20 degrees in ice), incoherent scattering extending across several Fresnel zones is
therefore likely
responsible for much of the signal extension observed. 
The assumption of incoherent scattering from macroscopic rubble near
the bed is also consistent
with our observation that signals observed with the Seavey antennas, which are responsive for frequencies greater than 250 MHz, are conspicuously more extended than the signals observed using the INR horns, which have good VSWR down to 100 MHz; longer-wavelength components should be less susceptible to decimeter-scaled bedrock scatterers \message{although this is not immediately borne out by a simple offline
high-pass filtering of the INR data!}
If the bottom 10 meters of the ice sheet also contains rubble, presumably
resulting from glacial motion across the bed, the resulting scattering
would also explain 
the apparent `pre-signal' observed in several of the waveforms.
The mixing of rock with basal water/ice was noted over 40 years ago
after the first core to bedrock was extracted at Byrd Station\citep{Gow68}.
\message{consistent with http://www.sciencemag.org/cgi/content/abstract/161/3845/1011 = ice core data to bedrock!}

\begin{figure}[htpb]
\centerline{\includegraphics[width=13cm]{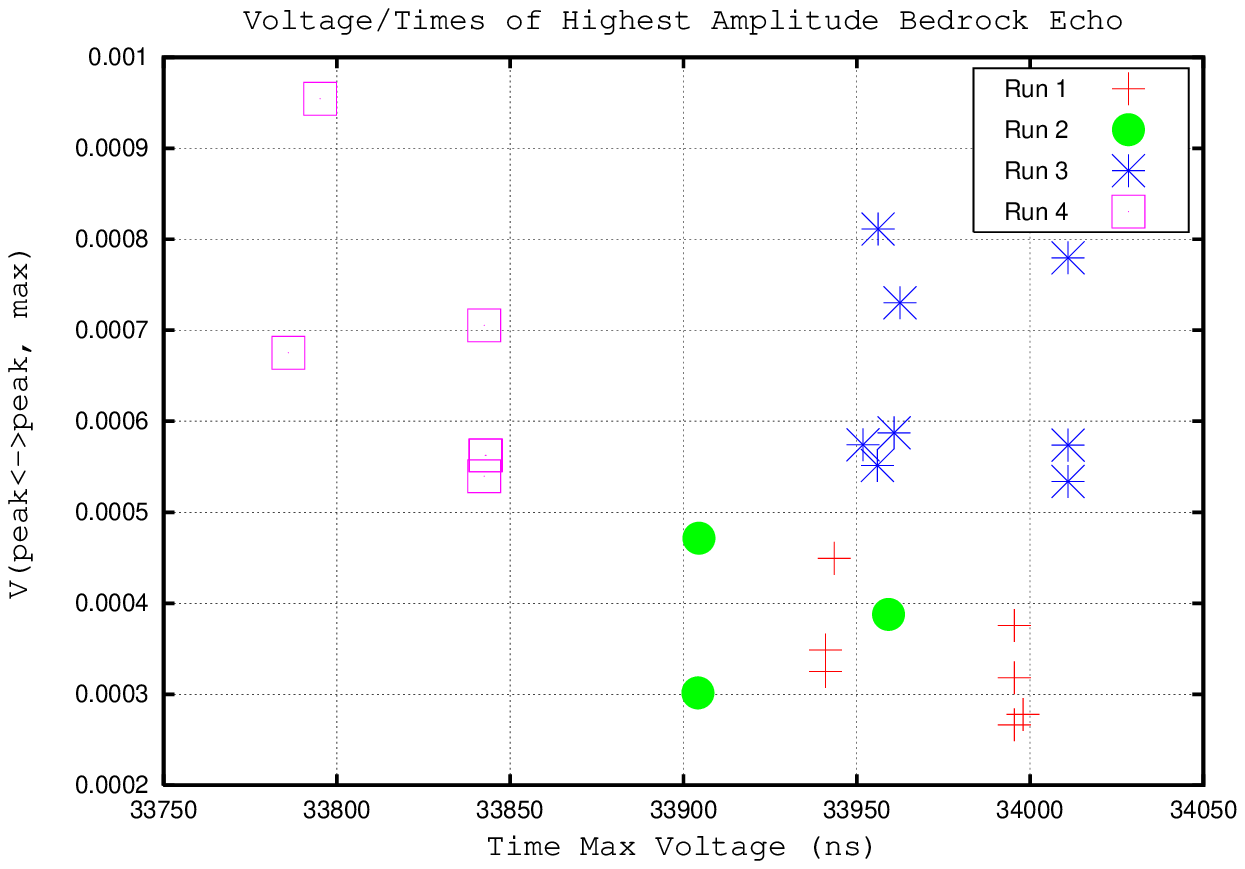}}
\caption{Time at which largest amplitude bedrock echo is registered in RICE
data acquisition system. Runs are distinguished as follows: Run 1: FID pulser,
 36 dB gain,
two Moscow INR horn antennas used as Tx/Rx, 100-1000 MHz bandpass; Run 2: same
as run 1, but with different cable lengths to antennas and antennas
displaced on snow surface; Run 3: FID pulser, 
52 dB gain 
(operating close to saturation), two Seavey antennas used as Tx/Rx, 
250-1000 MHz
bandpass; Run 4: HYPS pulser, 52 dB gain, Moscow INR horn antennas, 
150-1000 MHz
bandpass. Although the absolute echo times vary due to different run-to-run system delays, the time difference between the `early' vs. `late' times are consistently $\approx$50 ns.}
\label{fig:V1V2}
\end{figure}

\subsection*{Consistency of observed birefringent asymmetry}
Measurements were conducted using a variety of gains, as well as antennas. 
Due to varying locations of transmitter/receiver, as well as different instrumental
delays, the absolute time of the measured bottom echo can shift. Nevertheless,
for each separate `run' (defined as one stable hardware configuration),
when selecting the maximum voltage signal
observed in the interval 33000--35000 ns return time, 
a time delay of approximately 53 ns was observed in the signal arrival times
between the two modes (presumably corresponding to 
propagation along the major and minor birefringence axes), as we
rotate antennas in azimuth, as illustrated
in Figure \ref{fig:V1V2}.

\subsection*{Investigation of in-ice birefringence axis alignment}
\begin{figure}[htpb]
\centerline{\includegraphics[width=14cm]{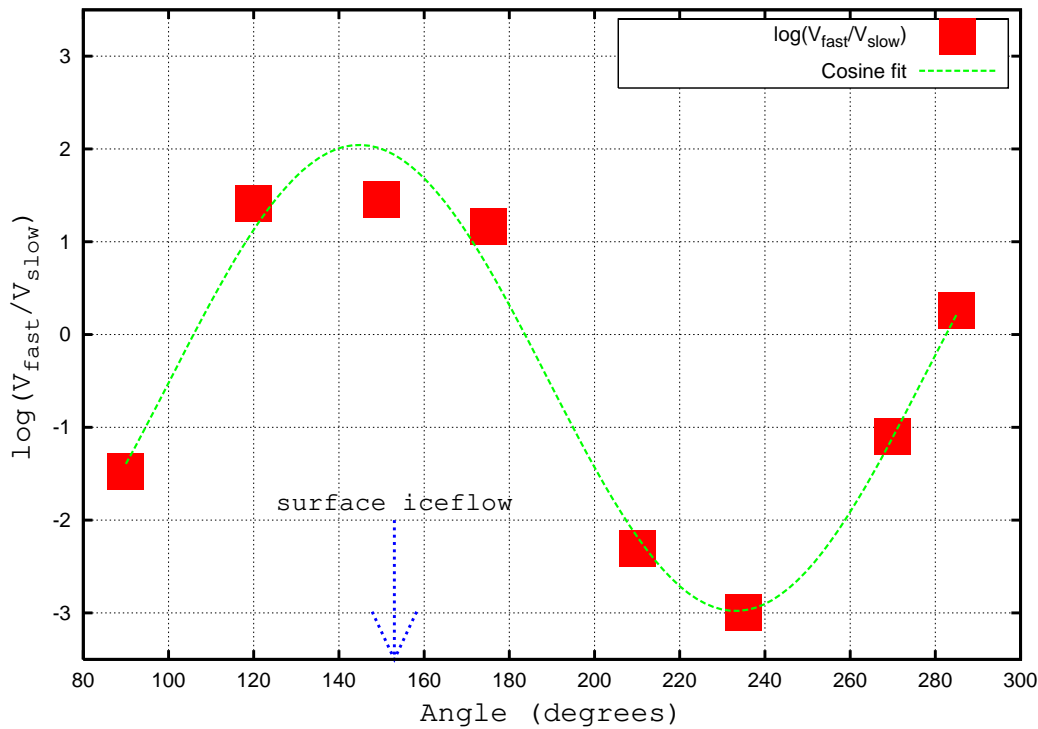}} 
\caption{Ratio of peak-to-peak voltage measured for
`early' bedrock reflection (``V1''), relative to peak-to-peak voltage measured
for `later' bedrock reflection (``V2''), after subtraction of rms noise.}
\label{fig:V1V2_1}
\end{figure}
When broadcasting along one of the two birefringent axes, we expect to observe only one
signal (assuming perfect isolation between the Vpol and Hpol terminals of our antennas);
when broadcasting at 45 degrees relative to the two birefringent axes, we expect to
observe equal amplitudes for the `early' vs. `late'
received signals. We have summarized
the relative amplitudes of the two signals in Figure \ref{fig:V1V2_1}; a value greater
than 1 indicates that the `early' signal was measured to have larger voltage (and 
vice versa). 
We note that this method of determining the optical axis orientation by the observed
elliptical polarization is very similar to the formalism outlined over three
decades ago by
Hargreaves\citep{Hargreaves-1977}. 
Interpreted as due exclusively to birefringence,
our data indicates that the birefringent axes are oriented at
approximately --40 and 50 degrees, respectively, relative to the zero degree axis, or within
25 degrees of the local ice flow direction.
Previous measurements of birefringence (\citep{Doake81,WoodruffDoake79}) 
found a lack of clear correlation between the optical axis and the 
local ice flow direction; those prior measurements attributed the
misalignment to a multi-step process leading to the in-ice crystal orientation.
We cannot, of course, exclude the possibility that an asymmetry in the 
imaginary portion of the dielectric constant is responsible for
at least some of the amplitude variation
we are observing; Fujita {\it et al}\citep{Fujita-1993} claimed an
extreme difference
of 15\% in the loss tangent for broadcasts parallel vs. perpendicular to
the optical axis.

\subsection*{Observed magnitudes of cross-polarized power}
\begin{figure} \centerline{\includegraphics[width=14cm]{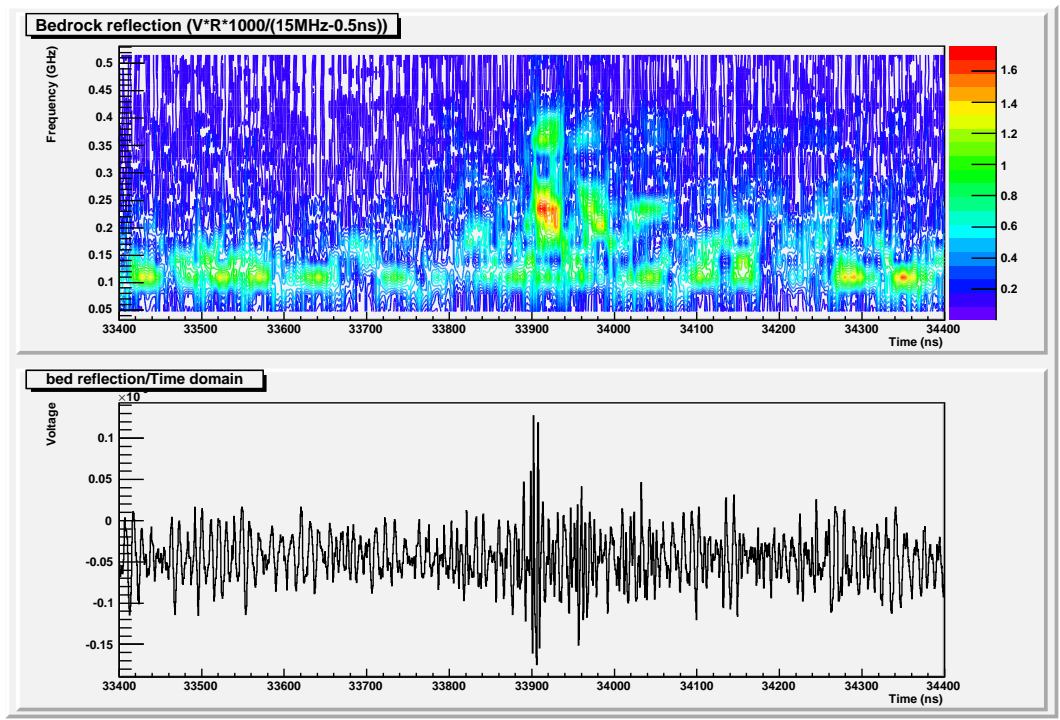}} \caption{Run 1 data, illustrating received signal strength in cross-polarization orientation - transmitter at +90 degrees; receiver at 0 degrees.}
\label{fig:XPol.eps} 
\end{figure}

In the absence of birefringence,
if we broadcast along one polarization, using antennas with 
excellent cross-polarization rejection, we would expect to observe
little or no received signal in a receiver with 
perpendicular alignment.
According to the manufacturer's
specifications for the Seavey antennas, the cross-polarization rejection
between is approximately 10:1 in voltage (better than 20 dB rejection
in power); we similarly
measure better than 14 dB rejection in the lab
for the INR antennas. Nevertheless, we note considerable 
cross-polarized power measured at our receivers, as shown in Figure \ref{fig:XPol.eps}.
This is, in fact, not the first observation of such co-polarization power at the Pole --
a reported rotation of the basal echo by 90 degrees was reported over 40 years ago\citep{Jiracek67}, 
consistent with our 2004 bedrock reflection data from South Pole\citep{RF-eps-im}.
For transmission at an angle $\theta$ relative to the ordinary axis, the signal amplitudes measured for the ``fast'' signal in the co-polarization vs. cross-polarization orientations should be $\cos^2\theta$ and $\cos\theta\sin\theta$, respectively. Conversely, the signal amplitudes, projected onto the extraordinary axis, in the co-polarization vs. cross-polarization orientations should be $\sin^2\theta$ and $\sin\theta\cos\theta$, respectively. If birefringence were solely responsible for the observed cross-polarized signal, we would therefore expect equal magnitudes of cross-polarized signal observed in all orientations, at variance with the observation in Figure \ref{fig:XPol.eps}. This suggests an additional
mechanism is at least partly responsible for the amplitude variation observed for the early vs. late bedrock echoes.
\message{See comment in TD.tex on this subject}

\begin{figure} \centerline{\includegraphics[width=13cm]{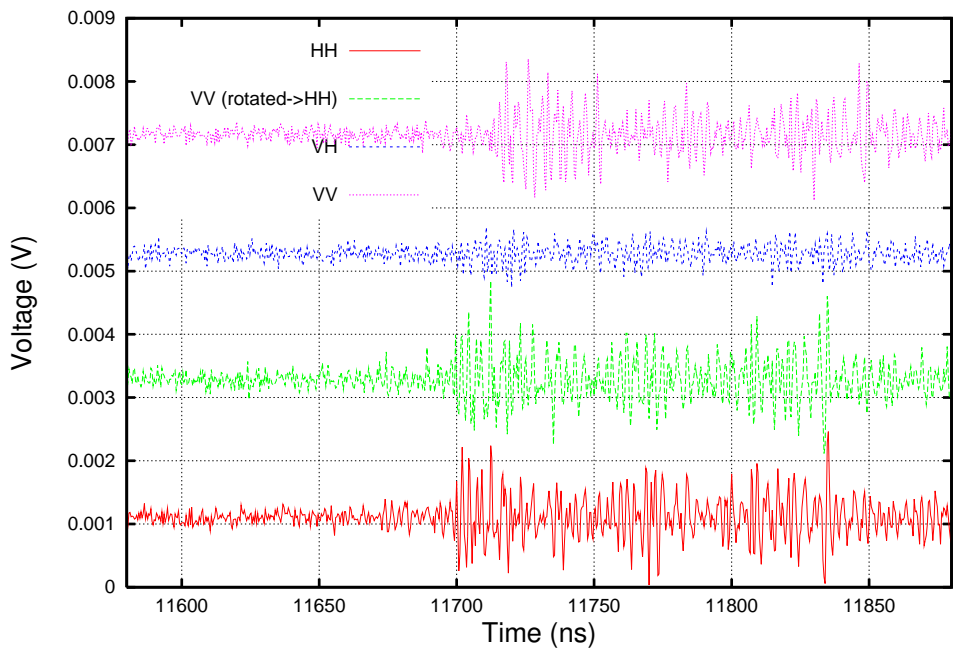}} \caption{2006 Taylor Dome
data, showing received signal as a function of time for indicated orientations.
``V'' and ``H'' designate orthogonal Vertical and Horizontal terminals of Seavey antennas, respectively.
``VV$\to$HH rotate'' refers to the Tx and Rx orientation for which
the VV axes of both Tx and Rx
have been rotated into the initial HH orientation.
In this Figure,
the three uppermost
signals have been vertically offset to 
enhance visual clarity.}
\label{fig:TD-HH-VV-1.eps} 
\end{figure}

Our 2008 analysis at South Pole\citep{SPRefl08} also showed considerable
observed cross-polarized signal strength for
internal reflections, in a regime where the average birefringent effect was found to be consistent with zero;
our measurements at Taylor Dome\citep{TD} displayed somewhat less cross-polarized
power in the bedrock reflection, as shown in Figure \ref{fig:TD-HH-VV-1.eps}.
Echoes provided by 
polarization-grating-like reflectors, as well as the mechanism
responsible for the observed extended bedrock echo signal shape may interfere
with birefringent asymmetries. The fact that
each of these has an associated, and generally unknown preferred axis
complicates an unambiguous association of the amplitude
data shown in Figure \ref{fig:V1V2_1} with the birefringent asymmetry
axis alone. If the reflecting plane is not horizontal, then there will also
be an asymmetry generated upon reflection, for the electric field components
perpendicular, and parallel to the reflecting surface, as well. 
In principle, Faraday rotation can also lead to considerable 
cross-polarized power;\footnote{Hargreaves excludes the possibility of
detectable Faraday rotation, based on the premise that a forward rotation
down to the base will be cancelled by a reverse rotation back to the 
surface. In fact, the phase inversion of the signal at the base results
in non-cancellation.} however, a recent laboratory study of cold ice
disfavors that possibility\citep{SPRefl08}.

\subsection*{Estimate of bedrock depth}
To determine the ice thickness at our measurement point,
we use the average time delay observed for run 1 (33972 ns), corrected
for the total cable delay (160.3 ns; Figure \ref{FID-output}) and use
a temperature-weighted average EM wave velocity of
$169\pm0.3$ m/$\mu$s below the firn\citep{Dowdeswell04}. 
Through the firn, we use direct measurements of radio propagation
wavespeed\citep{RF-eps-re},
implying a one-way transit time of 1102 ns through the upper 200 m of
the ice sheet at South Pole. 
Taking the errors from \citep{Dowdeswell04} gives an implied
depth of 
$2857\pm5$~m. 
However, field measurements of radio frequency
wave velocity have shown variations in the permittivity of up to 1\%; uncertainties
in impurity levels can, in principle, contribute an approximately equivalent
systematic error. Taken together, these imply a systematic depth error
of $\sim$30 m.

\subsection*{Amplitude Measurements - check of implied field attenuation length $L_{atten}$}
Knowing the total cable length
($\sim$125 m), the cable loss per unit
length ($\sim$5 dB/100 m at 500 MHz), the maximum signal
amplitude at the output of the pulser (2.5 kV), the measured
reflected signal amplitude $V_{ice}$, the bandpass of the antennas
($\sim$1 GHz), and the net gain of the amplifiers + filters
in the system, 
we can determine ``absolutely'' the   
average attenuation length to the bed by direct application of
the Friis Equation: $V_{Rx}/V_{Tx}=R\sqrt{G_{Tx}G_{Rx}}e^{-d_{tot}/L_{atten}}\lambda^2/(4\pi d_{tot})^2$, with
$R$ the bedrock reflectivity (taken to be 0.3), $V_{Rx}$ and $V_{Tx}$ the measured signal voltages, $G$ the antenna gain of transmitter and receiver antenna (taken to be 12 dBi given the expected in-ice focusing, and consistent with values used for our previous 2004 analysis with the same antennas\citep{RF-eps-im}), $d_{tot}$ the total round-trip echo path (5.7 km), and $\lambda$ the broadcast wavelength. Since we are using sharp time-domain signals, rather than the limited-duration `tone's used for our previous, dedicated attenuation length study, we do not expect the same precision as attained in that previous study, and correspondingly use $\lambda$=1 meter as sufficient for this coarse check of the received signal amplitudes. Averaged over
the entire vertical chord, we
obtain an implied average
attenuation length $L_{atten}=$627 m, entirely consistent with those
2004 measurements at Pole.

\subsection*{Comment on dispersion}
\begin{figure} \centerline{\includegraphics[width=13cm]{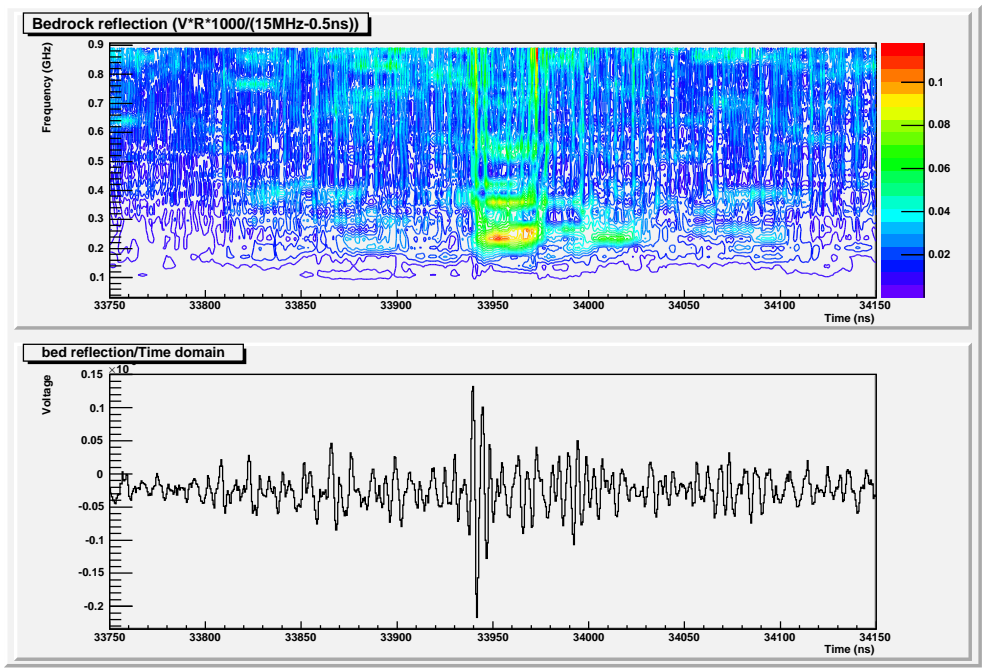}} \caption{Time domain and frequency domain (top) bedrock echo observed for
$-5^\circ$ orientation. We observe approximate synchronicity at all frequencies.} \label{fig:dispersion} \end{figure}

By performing a Fourier transform on the received waveforms, we can investigate
the dispersive characteristics of ice over the frequency range 200 MHz -- 900 MHz. Figure \ref{fig:dispersion} displays the Run 1 data at a Tx/Rx angle of --5 degrees. For this plot, the amplitude at high frequencies has been multiplied by frequency (squared) 
to compensate for the expected wavelength dependence of both receiver and transmitter
effective height. From this Figure, we estimate synchronicity of received power to within 4 ns for all frequencies considered, implying a variation in the real part of the dielectric constant less than 0.012\% over the range of frequencies considered in this analysis. This asymmetry is consistent with the
naive expectation that the variation in dielectric constant with frequency should be relatively small at radio frequencies, far from Debye resonances for ice. By contrast, in their extensive review of radio ice sheet sounding, Dowdeswell and Evans\citep{Dowdeswell04} suggest a variation in the dielectric constant of approximately $\Delta\epsilon'\approx0.04$ over the interval 1--100 MHz, corresponding to a wave speed variation of about 1.25\%.


\section*{Implications for Neutrino Detection}
An in-ice neutrino interaction can produce either a single
particle (muon) propagating over km-scale distances, or a compact
electromagnetic or hadronic shower, limited to $\sim$10 m in extent.
The former is best detected with suitably located photomultiplier tubes
sensitive to UV radiation. The latter is best detected with
englacial radio receivers sensitive to the expected $\sim$GHz bandwidth,
1-ns duration ``Askaryan'' signal impulse. Since the measured
birefringent asymmetry
is considerably larger in magnitude than 1 ns, a 1 GHz
bandwidth receiver will generally
observe `split' signals, with relative amplitudes
depending on the geometry and actual orientation
of the optical axis relative to the signal propagation direction. 
Monte Carlo simulations indicate that,
without compensating trigger electronics, the birefringent asymmetry measured
here would result in an approximately 5\% reduction
in equivalent neutrino detection volume. The smallness of this effect is
primarily due to the fact that a) the measured asymmetry is limited to deeper,
poorer quality (from the standpoint of RF transmission) Polar ice, b) the
measured asymmetry is restricted to the vertical axis, while neutrino signals
are predominantly observed along near-horizontal incident angles.

\section*{Summary and plan for additional work}
We have measured a birefringent effect, in the lower half of the ice
sheet at South Pole. The magnitude of the observed effect ($\sim$0.3\%)
is approximately consistent with other 
measurements of the radio frequency properties
of cold polar ice. It would be desirable to correlate our measurement with
direct physical measurements of South Polar ice down to the bed.
Unfortunately,
despite having perhaps the most developed infrastructural base in the
Antarctic interior, an ice core of the type taken at Siple Dome,
Dome C, Monning Draud or
Vostok has not yet been extracted. Such a core would provide incontrovertible,
and direct evidence for the COF correlation only suggested by our
current measurements. If such an association could be verified, the
application of birefringent and/or COF 
measurements to perform glacial tomography could
potentially develop as a discipline unto its own\citep{Eisen07,Eisen06}.
The fact that our measurements for the
upper half of the ice sheet are in direct contradiction to those
made at Dome Fuji imply that the 
ice characteristics are considerably different between those two locales;
the fact that the interior is often regarded as one
monolithic glacial mass notwithstanding.
Our results are, however, in agreement with the analysis of data taken in the
vicinity of Vostok, which found evidence for COF scattering dominating only
in the lowest third of the ice sheet, in agreement with direct core data\citep{Siegert00}.

\begin{figure} \centerline{\includegraphics[width=13cm]{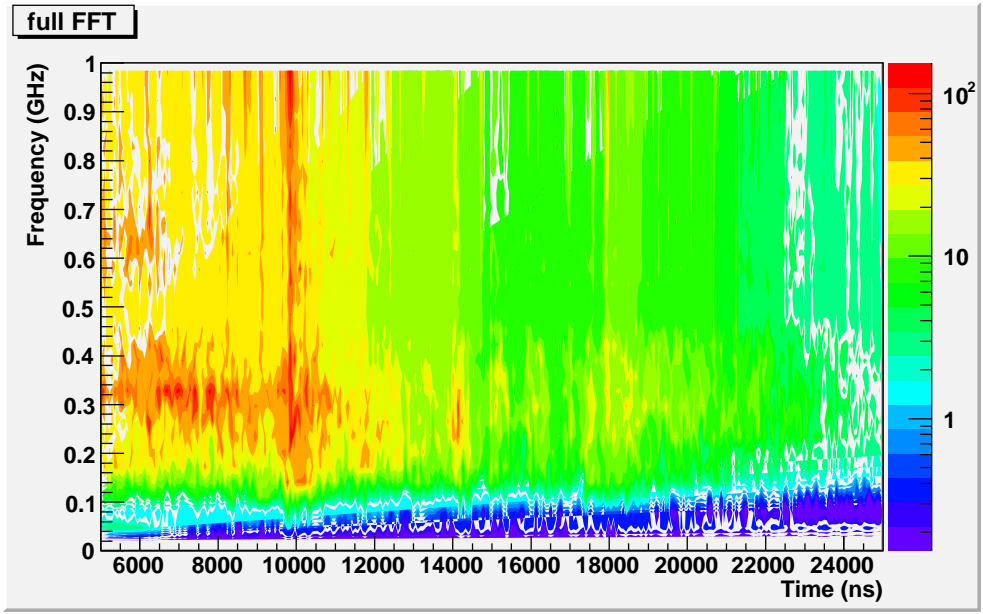}}\caption{Frequency domain amplitude observed for $0^\circ$ orientation, as a function of depth. Voltages have been scaled by $(frequency)^2$ to compensate for transmitter/receiver frequency response.} \label{fig:FullFFT} \end{figure}

A follow-up measurement would accumulate more cross-polarization data,
with yet a higher-power pulser, and attempt to
further elucidate the birefringent dependence on depth.
As Hargreaves\citep{Hargreaves-1977} points out, rotation of transmitter/receiver antennas through
the full azimuth in the cross-polarization orientation vs. rotation of Tx/Rx in the
co-polarization orientation allows one to separate out effects due to birefringence
vs. effects due to asymmetric reflecting planes. Assuming different
reflection coefficients in the x- vs. y-directions ($R_x$ and $R_y$), the maximum signal
measured with transmitter perpendicular to the receiver, compared to the minimum
signal measured in the same 
configuration allows an extraction of $R_x$ vs.
$R_y$. The matrix-based approach of
Fujita {\it et al.}\citep{Fujita06} also allows
a separation of the various contributions to anisotropic 
scattering based on a comprehensive analysis of the elliptical polarization 
characteristics
over a broad range of geometries and depths. 
Time limitations, however, prevented us from making the requisite suite of
measurements needed to fully exploit that formalism.
In addition to accumulating a more complete data sample,
a successor experiment would also take advantage of the ability to
broadcast, and study the reflections from, 
elliptically polarized signal of arbitrary ellipticity using the dual polarization ports of the Seavey
horn antennas. Also
particularly interesting would be data which might
further reveal the frequency dependence of echo returns. 
Figure \ref{fig:FullFFT}
displays the FFT, as a function of return time, for the $0^\circ$ orientation (chosen at random).
Although inconclusive, we observe considerable high-frequency power associated with, e.g., the return at $\sim$10000 ns, consistent with expectations from density/COF scattering. 
A pure conducting plane reflector would be expected to be: a) 
uniform as a function of rotation in azimuth, and b) have a power spectrum falling inversely
with frequency, at variance with our observations.
With enough resolution, one might hope to observe the sequence of
frequency-independent, frequency-dependent, and again frequency-independent
returns (and over the full azimuth)
that would be expected in a model where the scattering is
dominated by density, acid, and COF effects, respectively.
Additional data at higher power, and using a short-duration `tone' would be very useful in this respect.

\begin{figure}[htpb]
\begin{minipage}{18pc}
\centerline{\includegraphics[width=8cm]{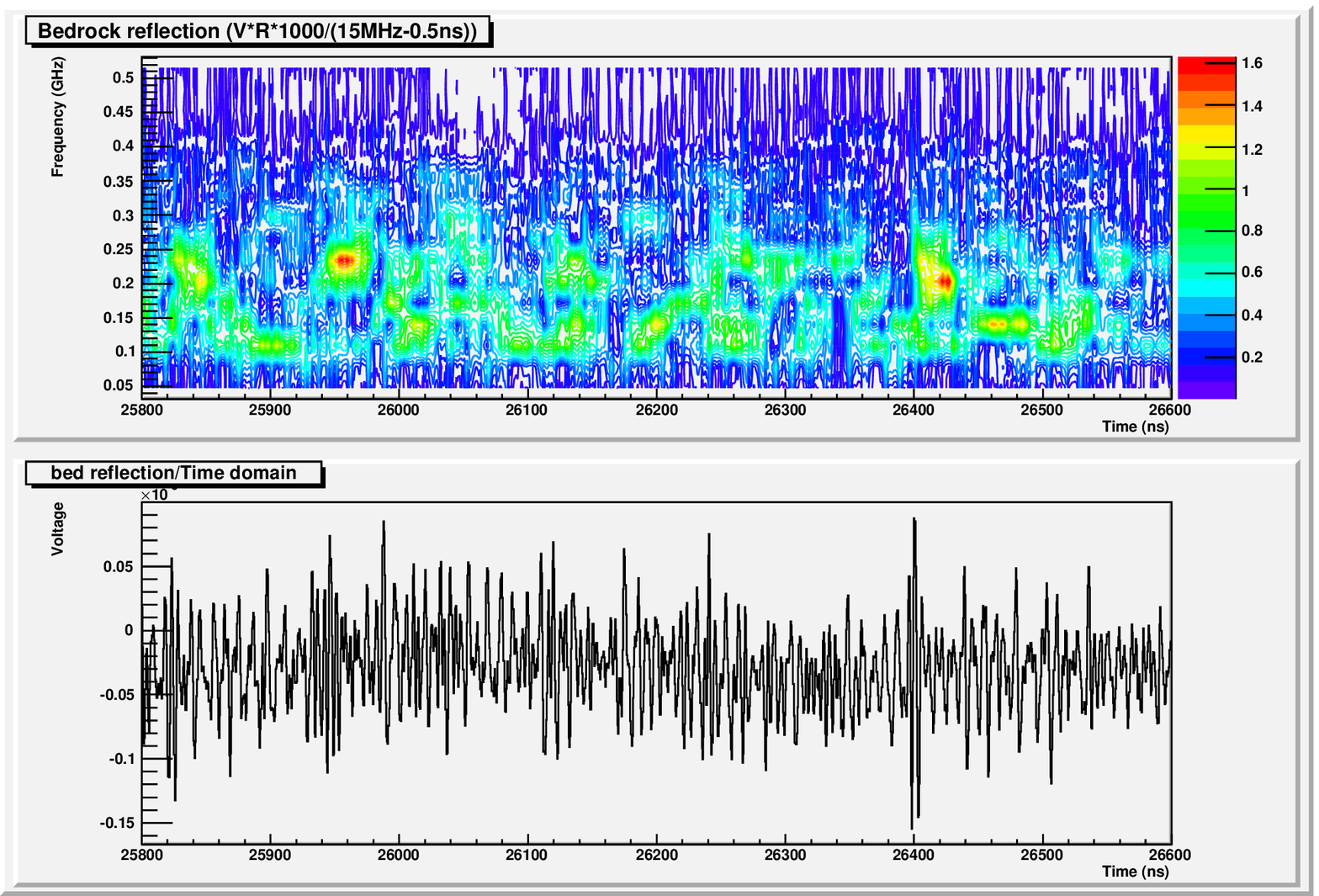}}
\caption{--10 x 0 Transmitter/Receiver orientation data, illustrating presence of possible return at 26.4 microseconds.}
\label{fig:25800-26600_0.eps}
\end{minipage}
\hspace{0.2pc}
\begin{minipage}{18pc}
\centerline{\includegraphics[width=8cm]{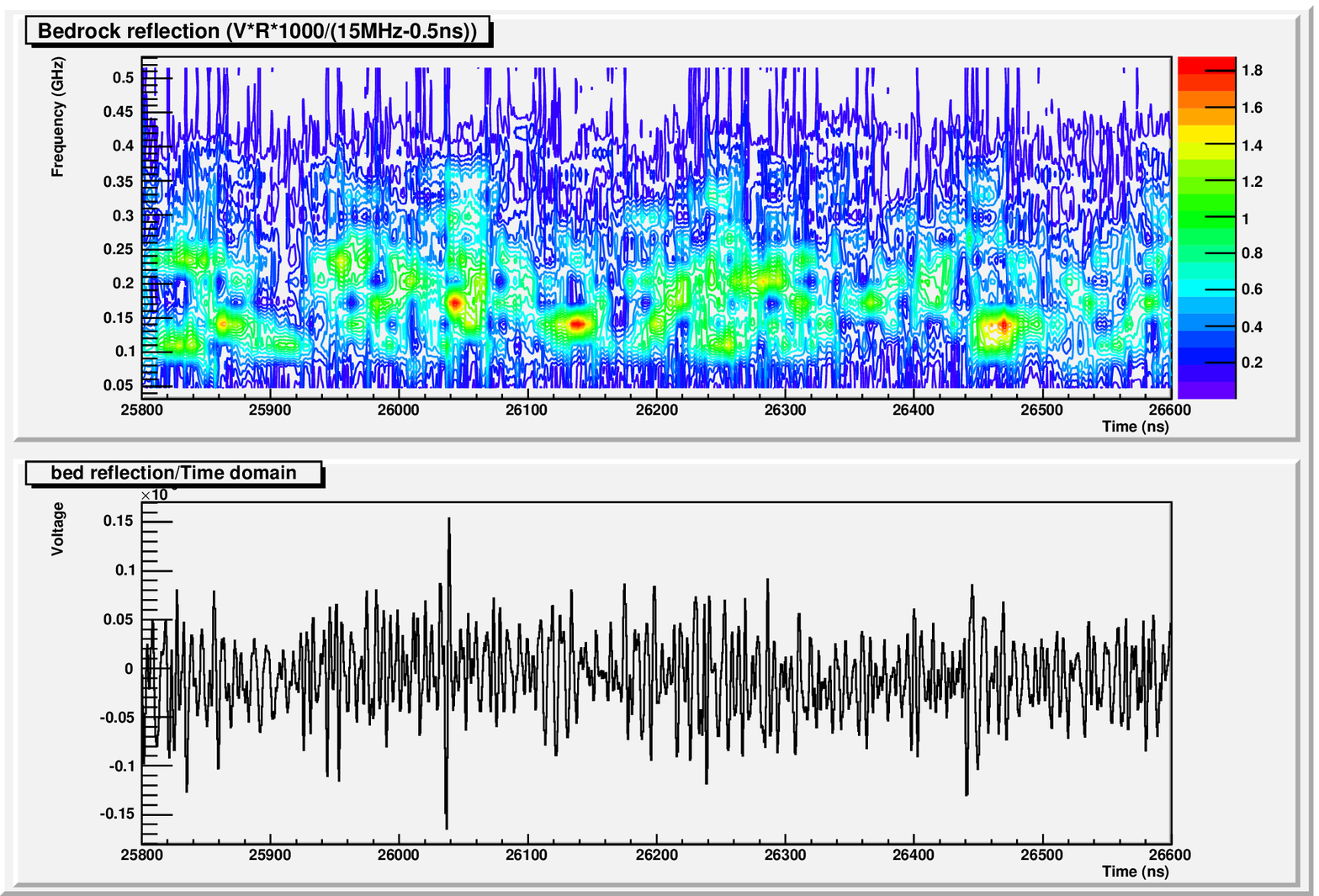}}
\caption{30 x 30 Tx/Rx orientation; Although 26.4 microsecond return is absent, note presence of apparent return at 26.04 microseconds.}
\label{fig:25800-26600_1.eps}
\end{minipage}
\end{figure}

\begin{figure}[htpb]
\vspace{-0.8cm}
\begin{minipage}{18pc}
\centerline{\includegraphics[width=8cm]{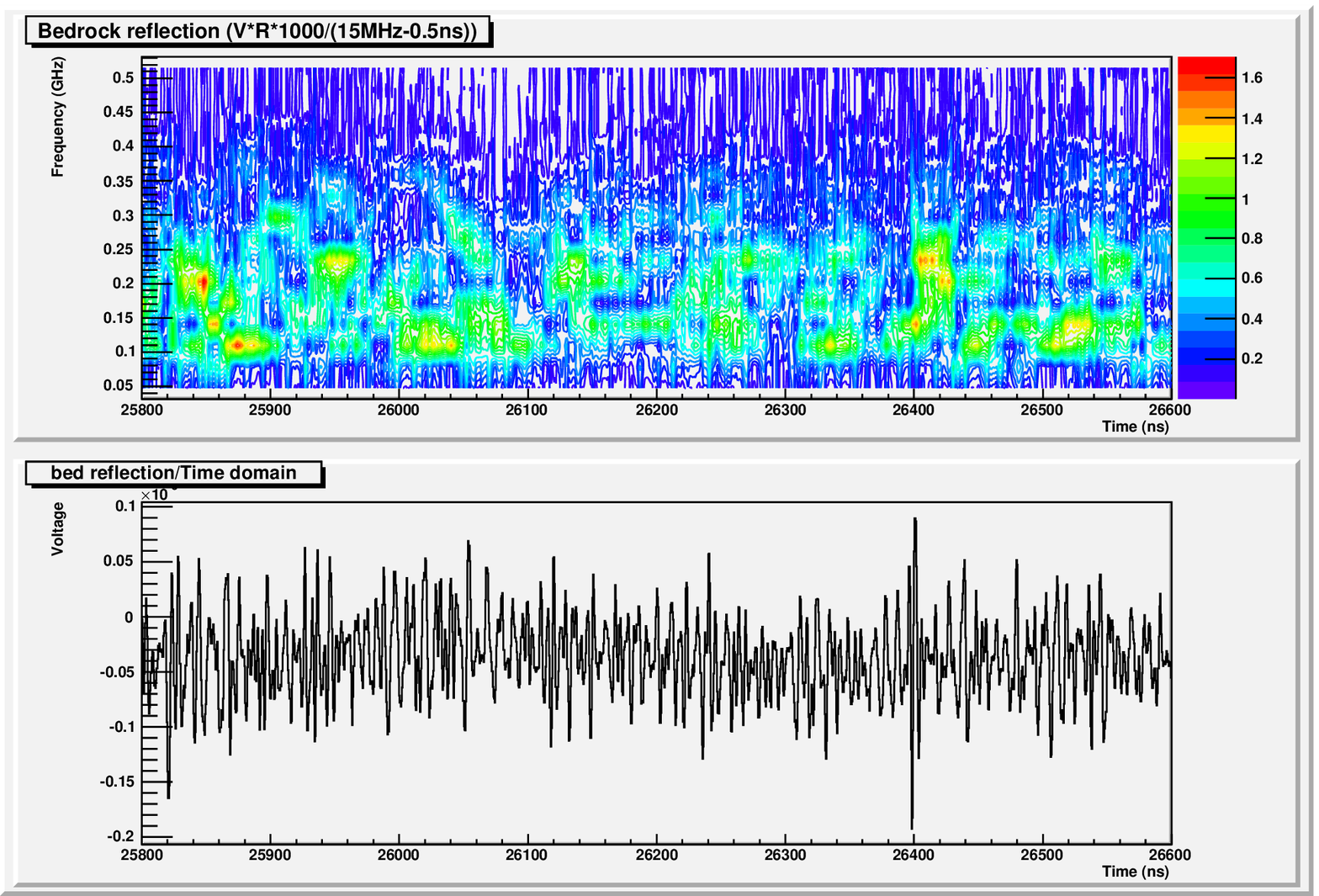}}
\caption{--30 x --30 orientation data.}
\label{fig:25800-26600_2.eps}
\end{minipage}
\hspace{0.2pc}
\begin{minipage}{18pc}
\centerline{\includegraphics[width=8cm]{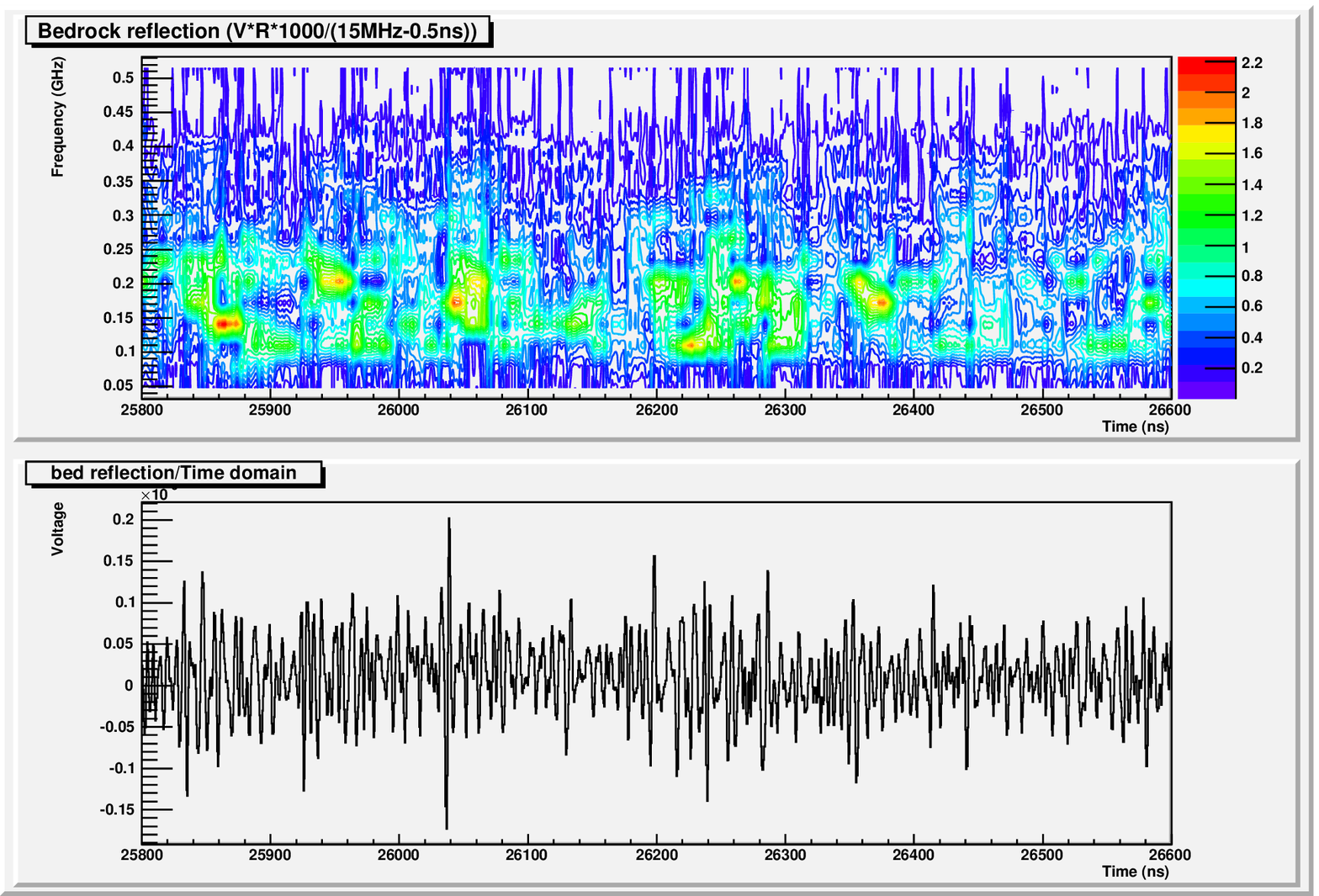}}
\caption{50 x 60 orientation data.}
\label{fig:25800-26600_3.eps}
\end{minipage}
\end{figure}

In the process of performing this measurement, one conspicuous
pattern that emerges is the similarity of waveforms delineated by
the optical axes we identify in this analysis. I.e., beyond
15000 ns, sharp
scattering lines consistently appear at the same depth for data
taken at angles $20^0-90^0$; a different set of 
reflectors are
evident for data taken at angles $-60^0-10^0$ (and consistently
within that angular interval). Similarly, signal shapes
(sharp or extended) of internally reflected radio waves
are repeatable within the two distinct angular ranges
(Figs \ref{fig:25800-26600_0.eps}, \ref{fig:25800-26600_1.eps}, \ref{fig:25800-26600_2.eps}, \ref{fig:25800-26600_3.eps});
this is
perhaps an indication that preferred crystal orientation is also accompanied by
an asymmetry in absorption. 

Finally, we
re-emphasize that we have sensitivity only to
the projection of the optical axis onto the vertical -- the true 
birefringent asymmetry could, in principle, be much larger
than the value reported here. The ARA experiment is expected to 
have the capability to broadcast in-ice signal
horizontally over several km and should therefore provide
essential vertical polarization information at radio frequencies.

\section*{Acknowledgments}
The authors particularly 
thank Chris Allen (U. of Kansas) and
Kenichi Matsuoka (U. of Washington) 
for very helpful discussions, as well as our colleagues on the
RICE and ANITA experiments. We also thank Andy Bricker of
Lawrence High School (Lawrence, KS) for his assistance working
with the Lawrence students. This work was supported by
the National Science Foundation's Office of Polar Programs
(grant OPP-0826747) and QuarkNet programs.
Any opinions, findings, and conclusions or 
recommendations expressed in this material are those of the author(s) 
and do not necessarily reflect the views of the National Science Foundation.

\end{document}